# Uncertainty-Aware Flow Field Reconstruction Using SVGP Kolmogorov-Arnold Networks

Y. Sungtaek Ju

Department of Mechanical and Aerospace Engineering, University of California, Los Angeles, CA 90095-1597, U.S.A.

## Abstract

Reconstructing time-resolved flow fields from temporally sparse velocimetry measurements is critical for characterizing many complex thermal–fluid systems. We introduce a machine-learning framework for uncertainty-aware flow reconstruction using sparse variational Gaussian processes in the Kolmogorov-Arnold network topology (SVGP-KAN). This approach extends the classical foundations of Linear Stochastic Estimation (LSE) and Spectral Analysis Modal Methods (SAMM) while enabling principled epistemic uncertainty quantification. We perform a systematic comparison of our framework with the classical reconstruction methods as well as Kalman filtering. Using synthetic data from pulsed impingement jet flows, we assess performance across fractional PIV sampling rates ranging from 0.5% to 10%. Evaluation metrics include reconstruction error, generalization gap, structure preservation, and uncertainty calibration. Our SVGP-KAN methods achieve reconstruction accuracy comparable to established methods, while also providing well-calibrated uncertainty estimates that reliably indicate when and where predictions degrade. The results demonstrate a robust, data-driven framework for flow field reconstruction with meaningful uncertainty quantification and offer practical guidance for experimental design in periodic flows.

## 1. Introduction

Full-field experimental characterization of unsteady flows remains challenging despite advances in optical diagnostics. Time-resolved particle image velocimetry (TR-PIV) at kilohertz rates is expensive and often impractical, while conventional PIV at lower repetition rates can miss important temporal dynamics. This measurement gap has motivated hybrid approaches that combine spatially rich but temporally sparse PIV data with temporally dense but spatially sparse sensor measurements [1–7]. Data-driven reconstruction methods have been effective in merging the spatial richness of low-repetition-rate PIV with the temporal resolution of point sensors.

The broad context of modal decomposition for flow analysis has been extensively reviewed [8]. Proper Orthogonal Decomposition (POD), introduced by Lumley [9] and developed extensively in subsequent studies [10,11], remains the foundation for data-driven flow analysis. Dynamic Mode Decomposition (DMD) [12,13] provides complementary time-domain insights, while SPOD [14] combines temporal and spatial spectral analysis. Gappy POD [15,16] addresses spatial reconstruction from incomplete data, while ensemble Kalman filtering [17] provides state estimation frameworks.

The time-domain Linear Stochastic Estimation (LSE) was first proposed [18] in the context of turbulent flow structures. This method builds a linear map (the estimator) that optimally relates the sensor measurements to the full flow field based on time-domain cross-correlations. This approach is often combined with Proper Orthogonal Decomposition (POD) to reduce the dimensionality of the problem, making the estimation more efficient and robust [19].

In recent years, extensions to this method have been developed in the frequency domain, such as the spectral LSE (SLSE) [20] and the Spectral Analysis Modal Methods (SAMMs) [21]. SAMMs reconstruct time-resolved flow fields by building upon SLSE approaches and Spectral POD (SPOD) [22]. They obtain dynamically important coherent structures from time-resolved local pressure measurements and non-time-resolved PIV measurements through a linear multi-input multi-output model. The method includes two variants: SAMM-SPOD and SAMM-RR (Rank Reduction). The reduced rank variant specifically addresses non-time-resolved PIV by computing cross-correlations using the full time-resolved sensor record shifted relative to each PIV snapshot. This enabled spectral estimation at arbitrary frequencies even with sparse PIV sampling.

State-space approaches offer an alternative framework through Kalman filtering [23,24], which combines a learned dynamics model with sensor observations to estimate the evolving flow state. The Kalman filter's recursive structure enables real-time implementation and provides uncertainty quantification through covariance propagation. However, as we learned in this work, the Kalman covariance matrix measures state uncertainty under model assumptions rather than prediction difficulty, limiting its utility for reliable uncertainty quantification at interpolation times.

Recent advances in scientific machine learning offer opportunities to address limitations of classical methods. Gaussian processes (GPs) provide a principled framework for nonlinear regression with native uncertainty quantification [25]. The sparse variational Gaussian process (SVGP) formulation [26,27] enables scaling to large datasets while preserving uncertainty estimates.

Recently developed Kolmogorov-Arnold Network (KAN) [28–30] is based on the Kolmogorov-Arnold representation theorem. It has received significant attention lately as an alternative to standard multi-layer perceptrons with improved interpretability and expressive power.

The combination of these two approaches in SVGP-KAN architectures [31,32] offers a path toward accurate, uncertainty-aware flow reconstruction. The SVGP-KAN architecture composes GPs within the KAN topology. It may be considered a variant of Deep Gaussian Processes [33] that imposes the additive structure of the Kolmogorov-Arnold theorem. This new neural network architecture is highly flexible and can be implemented into a wide variety of deep neural networks while maintaining the uncertainty quantification benefits of GP-based methods.

The present work provides a systematic comparison of established and new flow field reconstruction methods as applied to pulsed impingement jet flows, a configuration of practical importance for thermal management. Impinging jets are widely used or explored in various thermal management applications due to their high heat transfer coefficients [34,35]. Pulsed configurations can further enhance heat transfer through coherent vortex dynamics [36,37]. Accurate reconstruction of the unsteady flow fields is essential for understanding and optimizing these systems.

We demonstrate that SVGP-KAN and SAMM-RR methods achieve reconstruction accuracy comparable to that of the established methods across PIV sampling rates from 0.5–10%, with optimal performance at coprime sampling configurations that ensure uniform phase coverage. The key advantages of SVGP-KAN over SAMM-RR are twofold: (1) SVGP-KAN provides calibrated epistemic uncertainty estimates that correctly identify when and where predictions are less reliable, and (2) SVGP-KAN can capture nonlinear relationships between sensors and flow states through its composed GP architecture and thereby extends SAMM-RR, which constructs a single globally linear transfer function. We show that Kalman filter covariance, while theoretically motivated, has difficulty in providing meaningful uncertainty estimates at interpolation times due to a fundamental mismatch between what the covariance measures (state uncertainty under model assumptions) and what experimentalists need (expected prediction error). These findings provide practical guidelines for method selection in flow reconstruction applications.

## 2. Methods

We first discuss the basics of flow field reconstruction: Proper Orthogonal Decomposition (POD), Linear Stochastic Estimation (LSE), and Spectral Analysis Modal Methods (SAMM-RR). They form the foundational frameworks upon which our SVGP-KAN flow field reconstruction methods are constructed.

### 2.1 Proper Orthogonal Decomposition

The velocity field $\mathbf{u}(\mathbf{x}, t)$ is decomposed using POD as

$$\mathbf{u}(\mathbf{x}, t) = \bar{\mathbf{u}}(\mathbf{x}) + \sum_{k=1}^{N_m} a_k(t) \boldsymbol{\phi}_k(\mathbf{x})$$

where $\bar{\mathbf{u}}$ is the temporal mean, $\boldsymbol{\phi}_k$ are orthonormal spatial modes, $a_k(t)$ are time-dependent modal coefficients, and $N_m$ is the number of retained modes. The modes are computed via singular value decomposition of the fluctuating snapshot matrix $\mathbf{U}' = [\mathbf{u}'_1, \ldots, \mathbf{u}'_N]$, where $\mathbf{u}'_i = \mathbf{u}(\mathbf{x}, t_i) - \bar{\mathbf{u}}$ represents the fluctuation at PIV snapshot $i$.

For the present configuration with velocity components $(u, w)$, the state vector at each grid point is $[u', w']^T$, yielding a snapshot matrix of dimension $2N_x \times N_{PIV}$ where $N_x$ is the number of spatial grid points and $N_{PIV}$ is the number of PIV snapshots.

## 2.2 Linear Stochastic Estimation

The time-domain Linear Stochastic Estimation constructs a linear mapping between sensor fluctuation measurements $\mathbf{y}'(t) \in \mathbb{R}^{N_s}$ and modal coefficients $\mathbf{a}(t) \in \mathbb{R}^{N_m}$:

$$\hat{\mathbf{a}}(t) = \mathbf{C}\mathbf{y}'(t)$$

where $\mathbf{y}'(t) = \mathbf{y}(t) - \bar{\mathbf{y}}$ represents the fluctuating component of the sensor measurements. The transfer matrix $\mathbf{C} \in \mathbb{R}^{N_m \times N_s}$ is determined by minimizing the mean-squared estimation error at PIV times:

$$\mathbf{C} = \underset{\mathbf{C}}{\mathrm{argmin}} \sum_{i=1}^{N_{PIV}} \| \mathbf{a}(t_i) - \mathbf{C}\mathbf{y}'(t_i) \|^2$$

The solution is given by

$$\mathbf{C} = \mathbf{A}_{PIV} \mathbf{Y}'^{+}_{PIV}$$

where $\mathbf{A}_{PIV} = [\mathbf{a}(t_1), \ldots, \mathbf{a}(t_{N_{PIV}})]$ contains the true modal coefficients at PIV times computed by projecting PIV snapshots onto the POD basis. $\mathbf{Y}'_{PIV} = [\mathbf{y}'(t_1), \ldots, \mathbf{y}'(t_{N_{PIV}})]^T$ contains the corresponding sensor fluctuation measurements, and $(\cdot)^+$ denotes the Moore-Penrose pseudoinverse.

Two variants are considered: LSE (Vanilla) uses the sensor measurements directly, while LSE (Phase) augments the sensor inputs with sinusoidal phase features $[\sin(\phi), \cos(\phi), \sin(2\phi), \cos(2\phi)]$ where $\phi = 2\pi f_v t$ is the forcing phase.

## 2.3 Spectral Analysis Modal Methods (SAMM-RR)

The Spectral Analysis Modal Method (SAMM) operates in the frequency domain, constructing transfer functions based on cross-spectral densities. The sensor cross-spectral density matrix at frequency $f$ is

$$\mathbf{G}_{yy}(f) = \mathbb{E}[\hat{\mathbf{y}}(f)\hat{\mathbf{y}}^H(f)]$$

where $\hat{\mathbf{y}}(f)$ is the Fourier transform of the sensor signals and $(\cdot)^H$ denotes conjugate transpose. The cross-spectral density between sensors and modal coefficients is

$$\mathbf{G}_{ay}(f) = \mathbb{E}[\hat{\mathbf{a}}(f)\hat{\mathbf{y}}^H(f)]$$

The frequency-dependent transfer function is then

$$\mathbf{H}(f) = \mathbf{G}_{ay}(f)\mathbf{G}_{yy}^{-1}(f)$$

Zhang et al. [21] developed SAMM-RR specifically for non-time-resolved PIV combined with time-resolved sensors. Their key insight is that even though modal coefficients $\mathbf{a}(t)$ are available only at sparse PIV times, the cross-spectral density $\mathbf{G}_{ay}(f)$ can still be computed at arbitrary frequencies by exploiting the time-resolved sensor data. The cross-correlation is computed as:

$$R_{ay}(\tau) = \frac{1}{N_{PIV}} \sum_{i=1}^{N_{PIV}} \mathbf{a}(t_i)\mathbf{y}(t_i - \tau)^T$$

where $\mathbf{y}(t_i - \tau)$ is obtained from the full time-resolved sensor record by shifting relative to each PIV snapshot time. This enables computing $R_{ay}(\tau)$ at many time lags, and its FFT yields $\mathbf{G}_{ay}(f)$ at arbitrary frequencies.

To ensure consistent normalization in the transfer function, $\mathbf{G}_{yy}(f)$ is computed using the same correlogram approach applied to sensor data at PIV times:

$$R_{yy}(\tau) = \frac{1}{N_{PIV}} \sum_{i=1}^{N_{PIV}} \mathbf{y}(t_i)\mathbf{y}(t_i - \tau)^T$$

Both $\mathbf{G}_{ay}(f)$ and $\mathbf{G}_{yy}(f)$ are obtained by applying a window function to the correlation and computing the FFT, ensuring that the transfer function $\mathbf{H}(f)$ has correct magnitude scaling. The frequency-dependent eigen-structure of $\mathbf{G}_{yy}(f)$ is also exploited by SVGP KAN SAMM (Section 2.4.3) to derive a spectrally informed sensor basis for Gaussian process regression.

## 2.4 Sparse Variational Gaussian Process (SVGP) Kolmogorov-Arnold Networks (KAN)

### 2.4.1 Overview and Architecture

The SVGP-KAN approach replaces the linear sensor-to-mode mapping with a nonlinear Gaussian process model that provides calibrated uncertainty estimates. For each modal coefficient $a_k$, we model

$$a_k = f_k(\mathbf{x}) + \epsilon, \quad \epsilon \sim \mathcal{N}(0, \sigma_n^2)$$

where $f_k$ is a GP with the KAN-structured kernel and $\mathbf{x}$ is the augmented input vector containing sensor measurements and phase features.

Each edge function employs a Radial Basis Function (RBF) kernel:

$$k(x, x') = \sigma_f^2 \exp\left(-\frac{(x - x')^2}{2\ell^2}\right)$$

where $\sigma_f^2$ is the signal variance and $\ell$ is the lengthscale, both learned during training. The RBF kernel produces smooth reconstructions appropriate for the coherent vortex dynamics in our impingement jet case. Alternative kernels (e.g., Matérn) could be employed for flows with sharper gradients.

Rather than using periodic kernels, we capture the periodic nature of the pulsed jet through explicit phase features added to the input space. For time $t$ and forcing frequency $f_v$, we augment the sensor measurements with:

$$[\sin(\phi), \cos(\phi), \sin(2\phi), \cos(2\phi)]$$

where $\phi = 2\pi f_v t$ is the forcing phase. This approach offers several advantages. It separates the periodic structure from the kernel, maintaining the RBF kernel's smoothness properties, and the sinusoidal encoding avoids discontinuities at the $2\pi \to 0$ transition while capturing both fundamental and harmonic phase relationships.

The KAN architecture decomposes the mapping into layers of univariate functions:

$$f(\mathbf{x}) = \sum_{j=1}^{n_1} \varphi_{j,out}\left(\sum_{i=1}^{d} \varphi_{j,i}(x_i)\right)$$

where each edge function $\varphi_{j,i}$ is modeled as an independent SVGP. This structure provides interpretability while maintaining expressiveness through the composition of GPs.

*2.4.2 Sparse Variational Inference*

Exact GP inference requires $O(N^3)$ computation for matrix inversion, prohibiting application to large datasets. We employ sparse variational approximations [26,27] with $M$ inducing points $\mathbf{Z}$ and corresponding inducing variables $\mathbf{u}$ to reduce complexity from $O(N^3)$ to $O(NM^2)$ while preserving uncertainty quantification.

The variational posterior $q(\mathbf{u}) = \mathcal{N}(\mathbf{m}, \mathbf{S})$ approximates the true posterior, and the evidence lower bound (ELBO) is maximized:

$$\mathcal{L} = \mathbb{E}_{q(\mathbf{f})}[\log p(\mathbf{y}|\mathbf{f})] - \lambda \, \mathrm{KL}[q(\mathbf{u}) \,\|\, p(\mathbf{u})]$$

The predictive distribution at a new input $\mathbf{x}_*$ is Gaussian with mean and variance:

$$\mu_* = \mathbf{k}_*^T \mathbf{K}_{ZZ}^{-1} \mathbf{m}$$

$$\sigma_*^2 = k_{**} - \mathbf{k}_*^T \mathbf{K}_{ZZ}^{-1}(\mathbf{K}_{ZZ} - \mathbf{S})\mathbf{K}_{ZZ}^{-1}\mathbf{k}_*$$

where $\mathbf{k}_* = k(\mathbf{Z}, \mathbf{x}_*)$ and $k_{**} = k(\mathbf{x}_*, \mathbf{x}_*)$.

*2.4.3 Integration with LSE and SAMM Frameworks*

Two SVGP-KAN variants are implemented, differing in input preprocessing strategy: SVGP-KAN LSE and SVGP-KAN SAMM.

SVGP-KAN LSE replaces the linear transfer matrix $\mathbf{C}$ in standard LSE (Section 2.2) with a nonlinear GP-based mapping. Given sensor fluctuations $\mathbf{y}'(t) \in \mathbb{R}^{N_s}$ augmented with phase features $[\sin\phi, \cos\phi, \sin 2\phi, \cos 2\phi]$ where $\phi = 2\pi f_v t$, the model learns:

$$\hat{\mathbf{a}}(t) = f_{\text{SVGP-KAN}}([\mathbf{y}'(t); \sin\phi; \cos\phi; \sin 2\phi; \cos 2\phi])$$

The phase features encode the forcing cycle position, enabling the model to learn phase-dependent sensor-to-flow relationships that a purely linear mapping cannot capture.

SVGP-KAN SAMM incorporates spectral preprocessing derived from the SAMM-RR framework before the GP mapping. Rather than using simple time-domain covariance computed only at sparse PIV times, this variant exploits the full time-resolved sensor signal to capture frequency-dependent correlations.

The preprocessing proceeds as follows. First, the SAMM-RR spectral estimation (Section 2.3) is applied to compute the sensor cross-spectral density matrix $\mathbf{G}_{yy}(f) \in \mathbb{C}^{N_s \times N_s}$ at each frequency. The SVD of $\mathbf{G}_{yy}(f)$ yields frequency-dependent eigenvectors $\mathbf{U}(f)$ and singular values $\boldsymbol{\sigma}(f)$ representing the signal power distribution across sensor combinations at each frequency.

To obtain a single real-valued transformation matrix suitable for the GP input, we compute a power-weighted average across frequencies:

$$\bar{\mathbf{U}} = \sum_f w(f) \operatorname{Re}[\mathbf{U}(f)], \quad w(f) = \frac{\sum_i \sigma_i(f)}{\sum_{f'} \sum_i \sigma_i(f')}$$

where the weights $w(f)$ reflect the total sensor signal power at each frequency. This weighted average is then re-orthogonalized via SVD to yield an orthonormal basis $\mathbf{A}$. Sensor measurements are projected onto this spectrally informed basis:

$$\tilde{\mathbf{y}}(t) = \mathbf{A}^T \mathbf{y}'(t)$$

This spectral preprocessing captures frequency-dependent sensor correlations that time-domain covariance at sparse PIV times cannot resolve. For periodic flows dominated by the forcing frequency and its harmonics, the weighting naturally emphasizes the sensor combinations most relevant to the coherent dynamics. The GP then learns:

$$\hat{\mathbf{a}}(t) = f_{\text{SVGP-KAN}}([\tilde{\mathbf{y}}(t); \sin\phi; \cos\phi; \sin 2\phi; \cos 2\phi])$$

Both variants are trained on PIV snapshots. Modal coefficients $\mathbf{a}(t_i)$, computed by projecting PIV velocity fields onto the POD basis, serve as training targets. The ELBO objective is maximized using Adam optimization with minibatching over PIV samples. Our implementation uses a $[N_{in}, 16, N_m]$ network architecture where $N_{in} = N_s + 4$ accounts for sensors plus phase features, $M = 50$ inducing points per edge GP, KL divergence weight $\lambda = 0.001$, and edge sparsity weight 0.01. Training uses 500 epochs with learning rate 0.01. Following training, field-level calibration (Section 2.6) is applied using the PIV training data to ensure well-calibrated uncertainty estimates.

At test time, the trained model predicts modal coefficients $\hat{\mathbf{a}}(t)$ and associated uncertainties $\sigma_{\mathbf{a}}(t)$ at arbitrary sensor sampling times. The predictive variances from individual edge GPs combine through the KAN additive structure to yield modal uncertainties. These calibrated modal uncertainties then propagate to velocity field uncertainties via the POD modes (Section 2.6).

## 2.5 Kalman Filter Methods

The Kalman filter provides state estimation through a predict-update cycle. We test a basic linear forward Kalman filter and a Rauch-Tung-Striebel (RTS) smoother [38].

Forward Filter - Predict:

$$\hat{\mathbf{a}}_{k|k-1} = \mathbf{A}\hat{\mathbf{a}}_{k-1|k-1}$$

$$\mathbf{P}_{k|k-1} = \mathbf{A}\mathbf{P}_{k-1|k-1}\mathbf{A}^T + \mathbf{Q}$$

Forward Filter - Update (at each timestep):

$$\mathbf{K}_k = \mathbf{P}_{k|k-1}\mathbf{C}^T(\mathbf{C}\mathbf{P}_{k|k-1}\mathbf{C}^T + \mathbf{R})^{-1}$$

$$\hat{\mathbf{a}}_{k|k} = \hat{\mathbf{a}}_{k|k-1} + \mathbf{K}_k(\mathbf{y}_k - \mathbf{C}\hat{\mathbf{a}}_{k|k-1})$$

$$\mathbf{P}_{k|k} = (\mathbf{I} - \mathbf{K}_k\mathbf{C})\mathbf{P}_{k|k-1}$$

where $\mathbf{C}$ is the observation matrix relating modal coefficients to sensor measurements. At PIV times, an additional correction strongly constrains the state to match the observed modal coefficients.

RTS Smoother - Backward Pass:

$$\mathbf{G}_k = \mathbf{P}_{k|k}\mathbf{A}^T\mathbf{P}_{k+1|k}^{-1}$$

$$\hat{\mathbf{a}}_{k|T} = \hat{\mathbf{a}}_{k|k} + \mathbf{G}_k(\hat{\mathbf{a}}_{k+1|T} - \hat{\mathbf{a}}_{k+1|k})$$

$$\mathbf{P}_{k|T} = \mathbf{P}_{k|k} + \mathbf{G}_k(\mathbf{P}_{k+1|T} - \mathbf{P}_{k+1|k})\mathbf{G}_k^T$$

PIV measurements are treated as fixed-point constraints during the backward smoothing pass, preserving the accurate state estimates at observation times.

The dynamics matrix $\mathbf{A}$ is learned from PIV-to-PIV transitions via least-squares regression on the modal amplitudes computed by projecting PIV snapshots onto the POD basis. The observation matrix $\mathbf{C}$ maps from modal coefficients to sensor measurements. Process noise $\mathbf{Q}$ and measurement noise $\mathbf{R}$ covariances are estimated from the residuals of the dynamics and observation models, respectively.

As we did not observe any significant differences in the reconstruction of our period flows between the two linear Kalman filters, all the results shown are from the smoother unless otherwise noted.

We integrate the Kalman filters into the flow field reconstruction in two ways, differing in sensor preprocessing: Kalman LSE uses raw sensor measurements, while Kalman SAMM transforms sensors into a spectrally informed basis derived from the SAMM-RR framework. Specifically, Kalman SAMM computes the frequency-dependent eigen-decomposition of the sensor cross-spectral density $\mathbf{G}_{yy}(f)$ and forms a power-weighted average across frequencies, capturing frequency-dependent sensor correlations that exploit the full time-resolved data. Both variants use the same dynamics model learned from PIV-derived modal amplitudes.

## 2.6 Uncertainty Propagation and Calibration

### 2.6.1 Uncertainty Propagation

SVGP-KAN provides predictive variances $\sigma_{a_k}^2$ for each modal coefficient. These modal uncertainties propagate to the velocity field through the POD expansion. For the velocity fluctuation at spatial location $\mathbf{x}$:

$$\sigma_u^2(\mathbf{x}) = \sum_{k=1}^{N_m} \phi_k^2(\mathbf{x}) \, \sigma_{a_k}^2$$

where $\phi_k(\mathbf{x})$ is the $k$-th POD mode evaluated at $\mathbf{x}$. This assumes independence between modal coefficient uncertainties, which is consistent with the SVGP-KAN architecture where each mode is predicted by a separate output head.

### 2.6.2 Field-Level Calibration

Raw "uncertainties" from the GPKAN methods and Kalman filters require calibration to ensure that predicted confidence intervals achieve their nominal coverage. We employ field-level calibration using only PIV training data, computed during model fitting before any evaluation on interpolation times.

At each PIV training snapshot $i$, we compute the field-level prediction error $e_i(\mathbf{x}) = |u_{true}(\mathbf{x}, t_i) - u_{recon}(\mathbf{x}, t_i)|$ and corresponding uncertainty $\sigma_u(\mathbf{x}, t_i)$ at sampled spatial locations. The calibration factor $\beta$ is determined via zero-intercept linear regression:

$$\beta = \frac{\sum_{i,\mathbf{x}} \sigma_u(\mathbf{x}, t_i) \, e_i(\mathbf{x})}{\sum_{i,\mathbf{x}} \sigma_u^2(\mathbf{x}, t_i)}$$

All uncertainties are then scaled by $\beta$: $\tilde{\sigma}_u = \beta \sigma_u$. This field-level approach ensures calibration at the same level where we evaluate (velocity field), using only training data.

*2.6.3 Calibration Assessment*

We assess calibration quality on held-out interpolation times using three metrics:

Calibration slope $\alpha$: The slope of actual error versus predicted uncertainty, computed via zero-intercept regression. Values $\alpha \approx 1$ indicate good calibration; $\alpha < 1$ indicates overconfidence; $\alpha > 1$ indicates underconfidence.

Uncertainty ratio $\rho = \bar{\sigma}_{between}/\bar{\sigma}_{PIV}$: The ratio of mean uncertainty between PIV times to uncertainty at PIV times. A ratio near unity indicates that uncertainty correctly reflects prediction difficulty across all times; large ratios indicate uncertainty inflation between observations.

Coverage: The fraction of true values falling within predicted confidence intervals at various levels (68%, 90%, 95%).

We note that the uncertainty quantified here is *epistemic* (model confidence given available data), not *aleatoric* (intrinsic flow variability). For the deterministic periodic flow considered here, aleatoric uncertainty is negligible by construction.

## 2.7 Numerical Simulation and Data Generation

*2.7.1 Flow Configuration*

We generate synthetic data from direct numerical simulation of a pulsed axisymmetric impinging jet. The simulation uses COMSOL Multiphysics 6.3 with the following parameters: nozzle diameter $D = 10$ mm, nozzle-to-plate distance $H = 4D$, mean jet velocity $U_m = 10$ m/s (Reynolds number $Re = 6700$), and domain radius $6D$. The simulation domain and relevant boundary conditions are shown in Figure 1.

Multi-frequency forcing is imposed at the inlet to emulate natural jet instabilities:

$$U_{inlet}(t) = U_m[1 + 0.15\sin(2\pi f_v t) + 0.075\sin(2\pi f_s t) + 0.025\sin(2\pi f_h t)]$$

where $f_v = 400$ Hz (varicose mode, St = 0.4), $f_s = 800$ Hz (sinuous mode, St = 0.8), and $f_h = 1000$ Hz (helical mode, St = 1.0). This forcing produces coherent vortex ring structures that dominate the flow dynamics.

Gaussian noise at 2% of field standard deviation is added to PIV snapshots and 1% noise to sensor measurements, mimicking measurement conditions. Additional experiments at elevated noise levels (20% PIV, 2% sensor) are used to assess robustness.

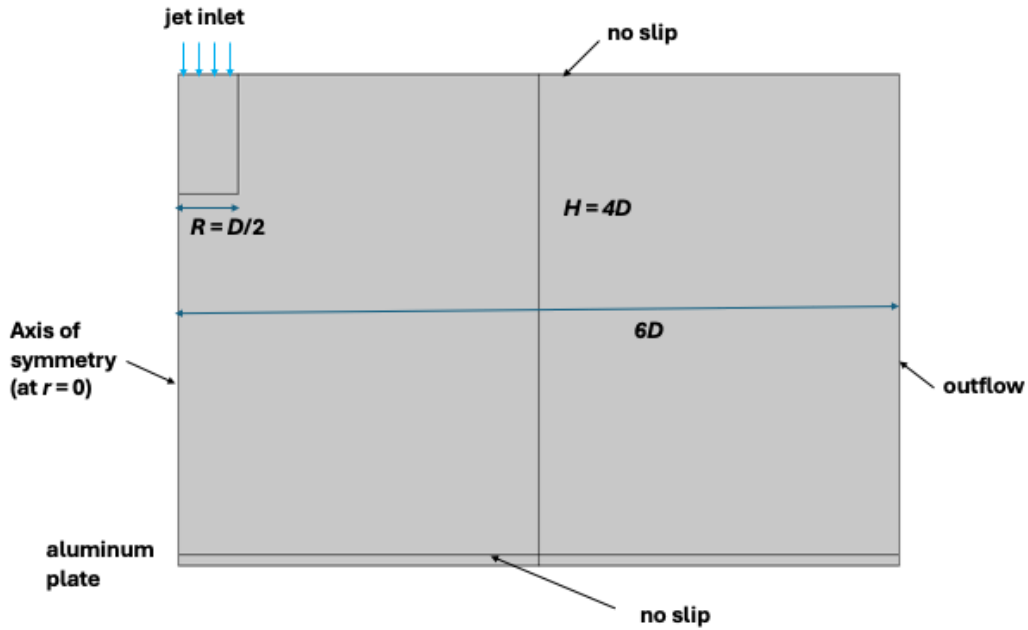

**Figure 1.** Numerical simulation domain and boundary conditions. Axisymmetric impinging jet with nozzle diameter $D = 10$ mm, nozzle-to-plate distance $H = 4D$, and domain radius $6D$.

Six pressure probes sample at 5 kHz: one near the nozzle exit and five distributed along the impingement plate at radial positions $r/D = 0.1, 0.5, 1.0, 2.0, 3.0$. This arrangement captures both free-jet and wall-jet dynamics while mimicking practical sensor constraints.

Seven fractional PIV sampling rates, measured with respect to the sensor sampling rate of 5 kHz, are examined (Table 1). The phase coverage varies substantially across configurations. For uniformly-spaced PIV samples, the number of distinct phases visited before the sequence repeats is $N_{phases} = N_{samples}/\gcd(N_{samples}, N_{cycles})$. Non-coprime configurations visit few distinct phases within the forcing cycle, but with many samples at each phase. Coprime configurations achieve near-uniform phase coverage with many distinct phases, but fewer samples per phase.

Table 1: PIV sampling configurations

| Sampling Rate | PIV Samples | Forcing Cycles | GCD | Distinct Phases | Samples/Phase | Coprime |
|---|---|---|---|---|---|---|
| 0.50% | 10 | 160 | 10 | 1 | 10.0 | No |
| 0.55% | 11 | 160 | 1 | 11 | 1.0 | Yes |
| 1.0% | 20 | 160 | 20 | 1 | 20.0 | No |
| 5.0% | 100 | 160 | 20 | 5 | 20.0 | No |
| 8.1% | 162 | 160 | 2 | 81 | 2.0 | Yes |
| 10.0% | 200 | 160 | 40 | 5 | 40.0 | No |
| 10.1% | 202 | 160 | 2 | 101 | 2.0 | Yes |

## 3. Results

### 3.1 POD Modal Analysis

Four POD modes are retained, capturing >99% of the fluctuating energy in the truncated modal subspace (Figure 2). The rapid energy convergence (Mode 1 capturing ~46%, Modes 1–2 capturing ~88%, and Modes 1–3 capturing ~94%) confirms the low-dimensional nature of the forced impingement jet flow, which is ideal for POD-based reconstruction methods.

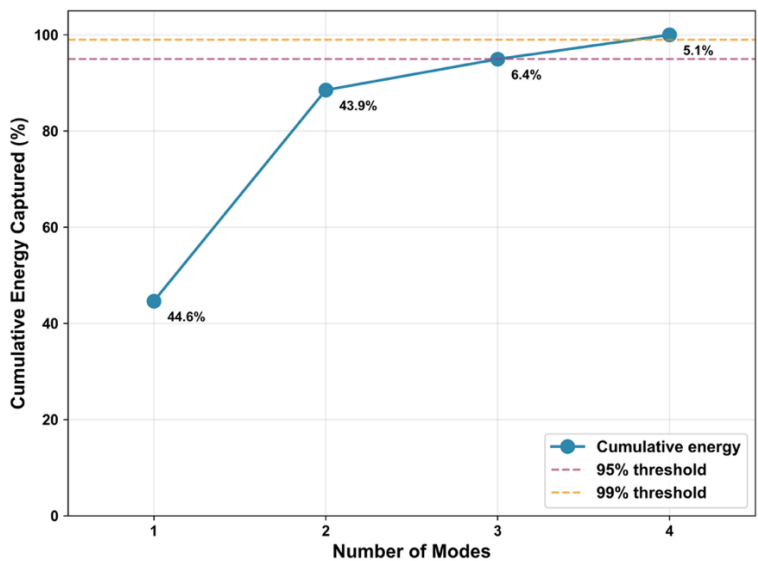

**Figure 2.** POD energy convergence. Cumulative energy captured versus number of modes showing rapid convergence: Mode 1 captures 45.6%, Modes 1-2 capture 87.7%, Modes 1-3 capture 94.1%, and Modes 1-4 capture 99.9% of fluctuating energy.

The spatial structure of the POD modes (Figure 3) reveals distinct flow physics corresponding to the imposed forcing frequencies. Mode 1 (45.6% energy) captures the dominant varicose dynamics at 400 Hz, with alternating positive/negative regions along the jet axis representing the axisymmetric pulsation. Mode 2 (42.1% energy), nearly equal in energy to Mode 1, represents a 90° phase-shifted version of the same physical mechanism. Modes 1 and 2 together form a traveling wave pair that captures vortex ring advection from nozzle to impingement surface. Mode 3 (6.4% energy) shows finer spatial structure concentrated near the impingement region, capturing the interaction between vortex rings and the wall jet. Mode 4 (5.8% energy) contains higher-order spatial features including contributions from the 800 Hz sinuous mode and 1000 Hz helical mode harmonics.

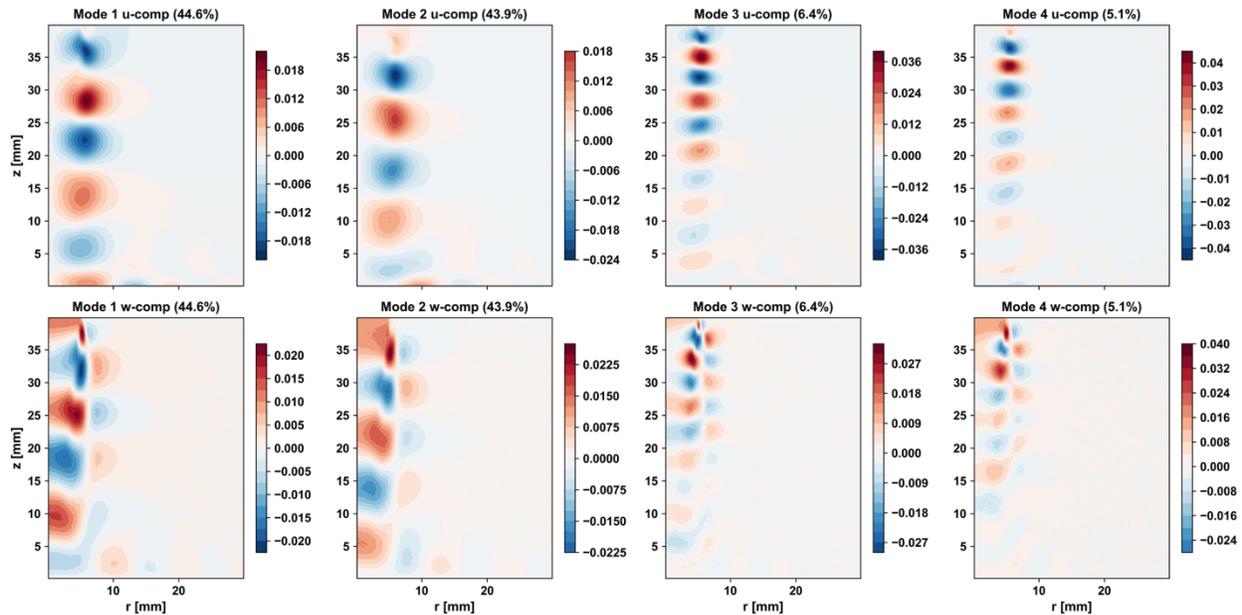

**Figure 3.** First four POD modes showing radial velocity $u$ (top row) and axial velocity $w$ (bottom row).

The spectral content of the pressure sensor signals (Figure 4) confirms the expected multi-frequency forcing. Figure 4(a) shows the average power spectral density across all sensors with the dominant peak at 400 Hz (varicose mode), along with clear secondary peaks at 800 Hz (sinuous) and harmonics at 200 Hz, 600 Hz, and 1000+ Hz arising from nonlinear interactions. Figure 4(b) displays individual sensor spectra, demonstrating consistent frequency content across measurement locations.

All six sensor channels detect the fundamental frequency, validating sensor placement and observability. The transfer matrix condition number $\kappa(\mathbf{C}) < 10^5$ across all sampling configurations indicates a reasonably well-posed inverse problem.

We note that the simulation operates in the deterministic, coherent-structure-dominated regime rather than the turbulent regime. At Re = 6700 with strong multi-frequency forcing (total amplitude ~20%), the flow response is dominated by phase-locked vortex dynamics rather than broadband turbulence. As discussed, four POD modes capturing >99% of fluctuating energy, whereas turbulent flows typically require 50–100+ modes. The discrete spectral peaks at forcing harmonics also do not exhibit broadband content (Figure 4). The SVGP-KAN homoscedastic likelihood assumption is therefore appropriate for this deterministic flow. Extension to turbulent flows with spatially varying aleatoric uncertainty would require heteroscedastic GP formulations.

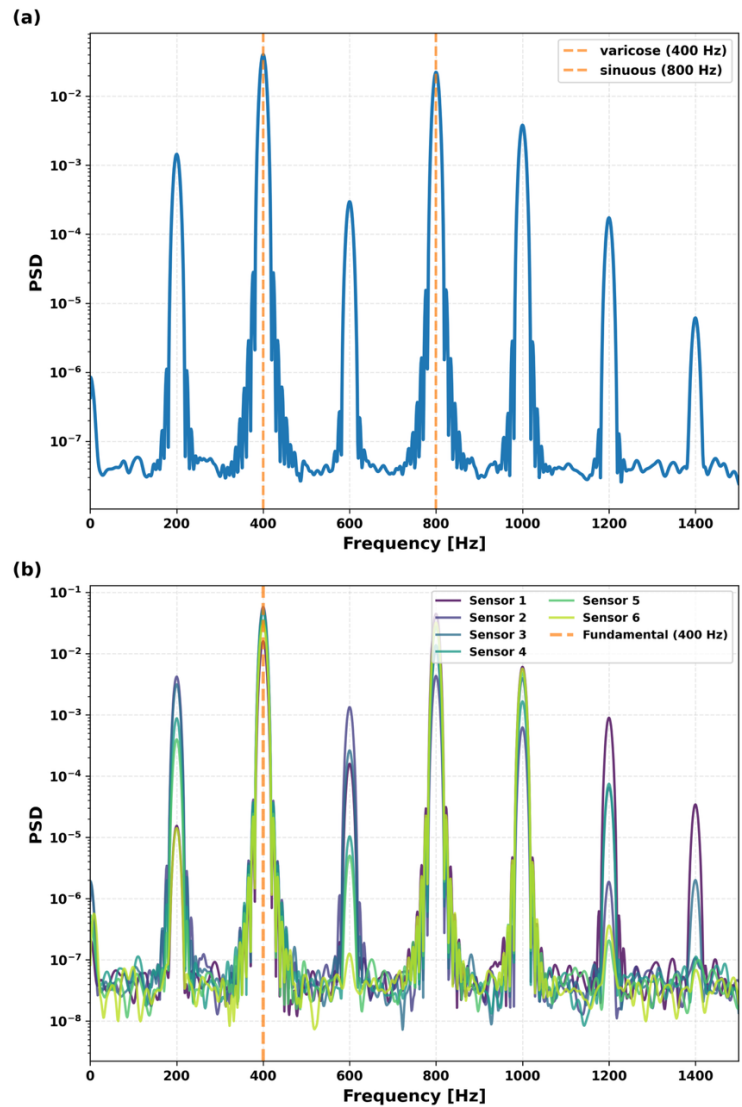

**Figure 4.** Spectral analysis of pressure sensor signals. (a) Average power spectral density showing dominant peak at 400 Hz (varicose mode) with harmonics. (b) Individual sensor spectra showing consistent frequency content across measurement locations.

## 3.2 Evaluation Metrics

We employ multiple metrics to assess reconstruction quality comprehensively.

L2 relative error quantifies global reconstruction accuracy:

$$\epsilon_{L2} = \frac{\| \mathbf{u}_{rec} - \mathbf{u}_{true} \|_2}{\| \mathbf{u}_{true} \|_2}$$

Generalization gap measures overfitting by comparing error at PIV times versus between-PIV times:

$$\Delta = \epsilon_{L2}^{interp} - \epsilon_{L2}^{PIV}$$

A large gap indicates the method fits PIV observations but fails to generalize.

Dynamic range ratio assesses amplitude preservation:

$$R_{dyn} = \frac{\text{std}(\mathbf{u}_{rec})}{\text{std}(\mathbf{u}_{true})}$$

Values below 1.0 indicate amplitude compression (smoothing); values above 1.0 indicate amplification.

Temporal correlation measures phase agreement:

$$r = \text{corr}(\mathbf{u}_{rec}(t), \mathbf{u}_{true}(t))$$

## 3.3 Reconstruction Accuracy

Table 2 presents the L2 interpolation error for all methods across key PIV sampling rates. Values represent errors at interpolation times (between PIV observations). SVGP-KAN and SAMM-RR methods achieve comparable accuracy, with the best performance at 8.1% coprime sampling. Kalman methods exhibit large errors at all sampling rates tested at interpolation times, with substantial generalization gaps between PIV and interpolation times. Third, at ultra-sparse rates (<1%), coprime configurations with only 1 sample per phase exhibit catastrophic failure while non-coprime configurations achieve substantially better results.

**Table 2: L2 interpolation error across PIV sampling rates (low noise: 2% PIV, 1% sensor)**

| Method | 0.55%† | 1.0% | 5.0% | 8.1%† | 10.0% |
|---|---|---|---|---|---|
| LSE (Vanilla) | 0.245 | 0.067 | 0.063 | 0.056 | 0.062 |
| LSE (Phase) | 0.245 | 0.084 | 0.081 | 0.061 | 0.069 |
| Kalman LSE | 0.627 | 0.116 | 0.112 | 0.087 | 0.138 |
| **SVGP-KAN LSE** | 0.237 | 0.068 | 0.054 | **0.047** | 0.054 |
| SAMM-RR | 0.245 | 0.072 | 0.054 | **0.047** | 0.054 |
| Kalman SAMM | 0.627 | 0.116 | 0.112 | 0.087 | 0.138 |
| **SVGP-KAN SAMM** | 0.237 | 0.070 | 0.054 | **0.047** | 0.054 |

*†Coprime sampling configuration. Best results in bold.*

The superior performance at 8.1% coprime sampling demonstrates the importance of phase coverage when sufficient samples per phase are available. With 81 distinct phases uniformly distributed across the forcing cycle and 2 samples per phase, the reconstruction methods can learn the full phase-dependent sensor-to-flow mapping. In contrast, ultra-sparse coprime configurations with only 1 sample per phase lack the redundancy needed for stable regression.

One notable finding is that Kalman-LSE and Kalman-SAMM produce identical results across all sampling rates and noise conditions. This equivalence is not a numerical coincidence but a mathematical necessity. The SAMM spectral basis **A** is derived from the eigen-decomposition of the sensor cross-spectral density matrix and is explicitly orthogonalized via SVD, yielding $\mathbf{A}^T\mathbf{A} = \mathbf{I}$.

For orthogonal transformations of the observation space, the Kalman filter state estimates remain invariant. Consider the transformed observations $\mathbf{w} = \mathbf{A}^T\mathbf{y}$. The observation model transforms as $\mathbf{C}_{new} = \mathbf{A}^T\mathbf{C}$, and the observation noise covariance as $\mathbf{R}_{new} = \mathbf{A}^T\mathbf{R}\mathbf{A}$. The Kalman gain becomes $\mathbf{K}_{new} = \mathbf{K}\mathbf{A}$, and the innovation $\tilde{\mathbf{w}} = \mathbf{A}^T\tilde{\mathbf{y}}$. The state update then yields:

$$\hat{\mathbf{x}} = \hat{\mathbf{x}}^- + \mathbf{K}_{new}\tilde{\mathbf{w}} = \hat{\mathbf{x}}^- + \mathbf{K}\mathbf{A}\mathbf{A}^T\tilde{\mathbf{y}} = \hat{\mathbf{x}}^- + \mathbf{K}\tilde{\mathbf{y}}$$

since $\mathbf{A}\mathbf{A}^T = \mathbf{I}$ for orthogonal matrices. Thus, orthogonal rotations of sensor space are mathematically invisible to the Kalman filter.

This result clarifies why SAMM spectral preprocessing benefits direct sensor-to-mode mapping methods (SVGP-KAN, SAMM-RR) but does not improve Kalman filtering. The learned observation matrix **C** already captures the optimal linear relationship between sensors and modal states; rotating the sensor basis merely rotates **C** correspondingly without changing the underlying estimation. In contrast, SVGP-KAN and SAMM-RR benefit from the spectrally informed basis because it decorrelates sensor inputs and emphasizes frequency bands with high signal-to-noise ratio, improving the conditioning of the nonlinear regression problem.

## 3.4 Generalization Gap

Table 3 shows the generalization gap (L2 at interpolation times minus L2 at PIV times) across sampling rates. The Kalman methods exhibit relatively large generalization gaps at all sampling rates, indicating that Kalman filtering achieves low error at PIV measurement times but fails to propagate information accurately between measurements. This is a symptom of dynamics model breakdown that is not resolved even at higher sampling rates. SVGP-KAN, SAMM-RR, and LSE methods maintain negligible gaps across all conditions, demonstrating generalization. See Figure 5 and Section 3.5 for further discussion of the phase coverage and generalization gap at low sampling rates.

Table 3: Generalization gap across sampling rates

| Sampling Rate | Steps (k) | LSE Gap | SVGP-KAN LSE Gap | Kalman LSE Gap | SAMM-RR Gap |
|---|---|---|---|---|---|
| 1.0% | ~100 | 0.007 | 0.017 | 0.064 | 0.008 |
| 5.0% | ~20 | 0.000 | 0.000 | 0.057 | 0.000 |
| 8.1% | ~12 | 0.000 | -0.001 | 0.038 | 0.000 |
| 10.0% | ~10 | 0.000 | 0.000 | 0.083 | 0.000 |

## 3.5 Structure Preservation

Table 4 summarizes structure preservation metrics. SVGP-KAN and SAMM-RR methods achieve the highest correlations across moderate sampling rates, with peak performance at 8.1% coprime sampling. Kalman methods show reduced correlation due to smoothing effects from dynamics model limitations.

Table 4: Structure preservation metrics at 5% and 8.1% sampling

| Method | Dynamic Range (5%) | Correlation (5%) | Dynamic Range (8.1%) | Correlation (8.1%) |
|---|---|---|---|---|
| LSE (Vanilla) | 0.96 | 0.992 | 0.96 | 0.994 |
| Kalman LSE | 0.90 | 0.978 | 0.99 | 0.983 |
| SVGP-KAN LSE | 0.97 | **0.993** | 0.97 | **0.996** |
| SAMM-RR | 0.97 | **0.993** | 0.98 | **0.996** |
| SVGP-KAN SAMM | 0.97 | **0.993** | 0.97 | **0.996** |

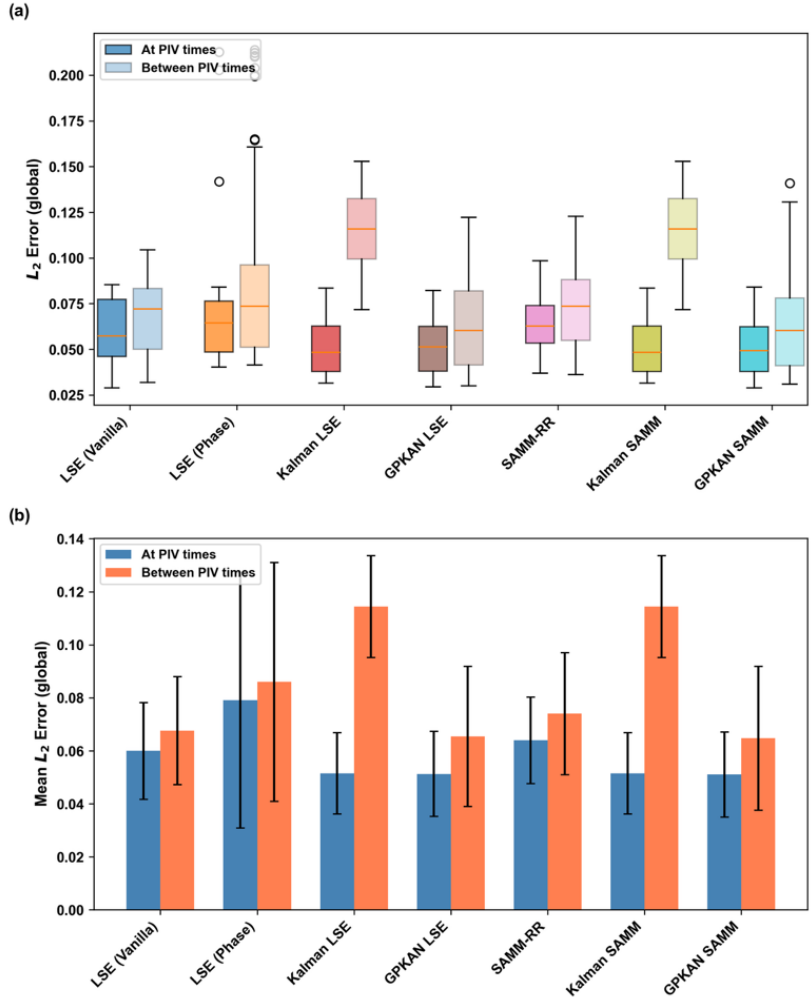

**Figure 5.** Generalization gap analysis at 1% non-coprime sampling. (a) Box plots showing L2 error distributions at PIV times (dark boxes) versus between PIV times (light boxes) for all reconstruction methods. (b) Mean L2 errors with standard deviations.

## 3.6 Visual Comparison

Figure 6 shows representative snapshot reconstructions for LSE-based methods at 5% PIV sampling rate. The true flow field exhibits distinct vortex structures propagating from the nozzle exit toward the impingement plate. SVGP-KAN LSE captures these structures with correct spatial localization and amplitude. LSE (Vanilla) provides good reconstruction with slightly reduced amplitude in vortex cores. Kalman LSE shows visible smoothing and amplitude reduction, particularly in the jet cores, with substantially higher L2 error.

Figure 7 shows the corresponding SAMM-based methods. SAMM-RR achieves accurate reconstruction comparable to SVGP-KAN methods. Kalman SAMM exhibits similar smoothing to Kalman LSE. SVGP-KAN SAMM recovers the vortical structures comparably to SVGP-KAN LSE.

Figure 8 shows vorticity evolution over one complete forcing period ($T$ = 2.5 ms at 400 Hz) at 5% PIV sampling rate. The top row displays SVGP-KAN LSE reconstructions at eight equally spaced phases, while the bottom row shows the corresponding ground truth fields. The reconstruction accurately captures the phase-resolved dynamics: vortex ring formation at the nozzle exit, advection through the free jet region, impingement on the plate surface, and radial spreading.

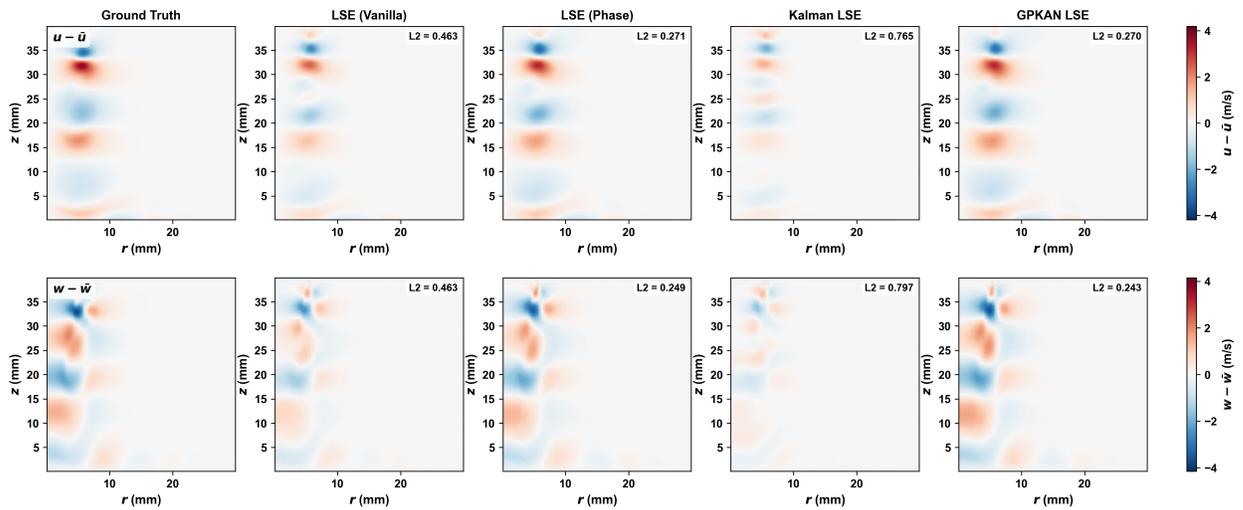

**Figure 6.** Snapshot reconstructions for LSE-based methods at 5% sampling. Left column: ground truth fluctuating velocity. Columns 2–5: reconstructions from LSE (Vanilla), LSE (Phase), Kalman LSE, and SVGP-KAN LSE. Top row: radial velocity $u - \bar{u}$. Bottom row: axial velocity $w - \bar{w}$. L2 errors shown in upper right of each panel.

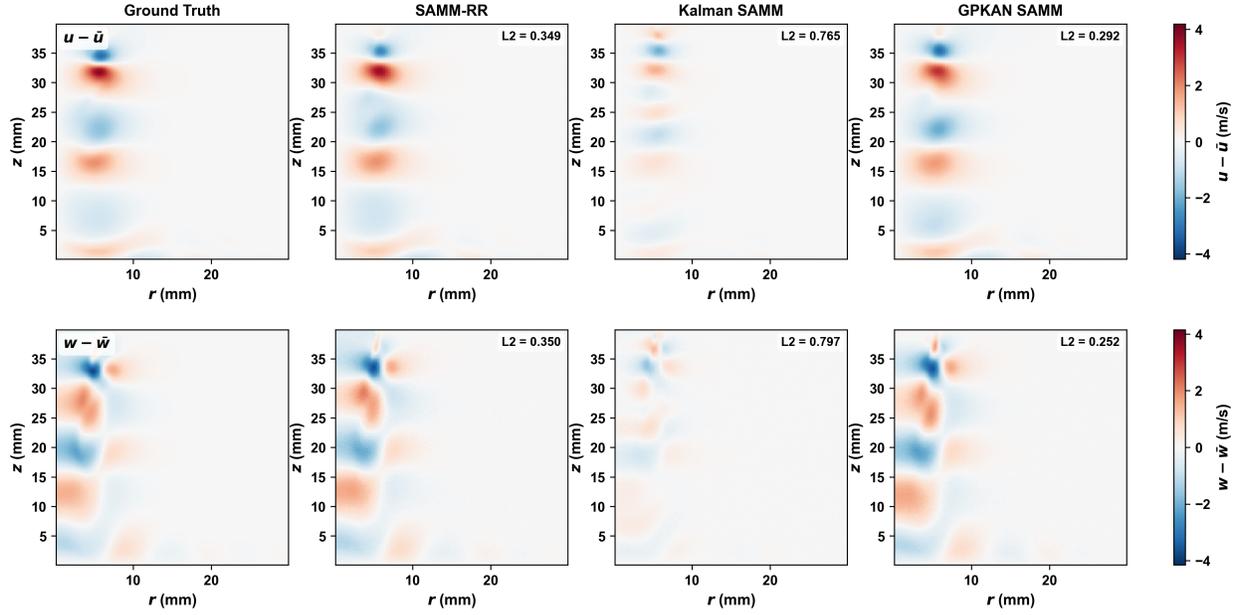

**Figure 7.** Snapshot reconstructions for SAMM-based methods at 5% sampling. Left column: ground truth. Columns 2–4: reconstructions from SAMM-RR, Kalman SAMM, and SVGP-KAN SAMM.

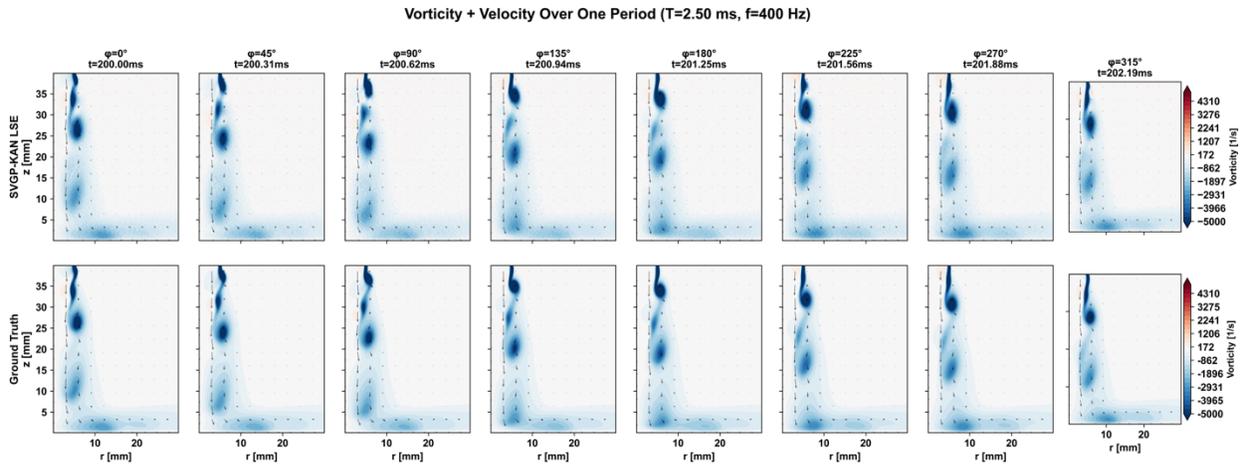

**Figure 8.** Vorticity evolution over one forcing period at 400 Hz (T = 2.5 ms) at 5% sampling. Top row: SVGP-KAN LSE reconstruction at eight phases (ϕ = 0°, 45°, ..., 315°). Bottom row: ground truth.

## 3.7 Ultra-Sparse Sampling: The 2 Samples/Phase Threshold

One additional finding emerges from the ultra-sparse sampling experiments is that coprime sampling configurations fail when fewer than 2 samples per phase are obtained. Table 5 highlights this threshold effect.

Figure 5 shows the phase coverage analysis results, visualizing the generalization gap for the different methods at 1% (non-coprime) sampling rate. Figure 5(a) shows box plots of L2 error distributions at PIV times versus between PIV times for all methods. Figure 5(b) presents mean errors with standard deviations. While Kalman methods achieve low error at PIV times (L2 ≈ 0.05), their performance degrades substantially at interpolation times (L2 ≈ 0.11), resulting in a generalization gap of ~0.06. In contrast, SVGP-KAN and SAMM-RR methods maintain consistent accuracy between PIV and interpolation times, demonstrating better generalization capability.

Table 5: Ultra-sparse sampling performance (SVGP-KAN LSE)

| Rate | Type | Phases | Samples/Phase | L2 Error | Correlation |
|---|---|---|---|---|---|
| 0.50% | Non-coprime | 1 | 10.0 | 0.089 | 0.984 |
| 0.55% | Coprime | 11 | 1.0 | 0.237 | 0.896 |
| 1.0% | Non-coprime | 1 | 20.0 | 0.068 | 0.990 |
| 8.1% | Coprime | 81 | 2.0 | **0.047** | **0.996** |

At ultra-sparse rates with 1 sample per phase (0.55%), coprime configurations exhibit large L2 errors, approximately 3-4x worse than non-coprime alternatives with concentrated sampling. This failure demonstrates that phase diversity without redundancy is insufficient for stable reconstruction.

The threshold of 2 samples per phase represents the minimum redundancy required for coprime sampling to outperform non-coprime alternatives. At and above this threshold (8.1%, 10.1%), coprime configurations achieve the best results by combining phase diversity with sufficient per-phase sampling.

## 3.8 Uncertainty Quantification

Figure 9 presents calibration curves for the four uncertainty-quantifying methods at 5% sampling rate. Table 6 summarizes the calibration slopes and related metrics across multiple sampling rates.

**Table 6: Uncertainty calibration metrics at interpolation times**

| Sampling | Method | Slope | $R^2$ | σ Ratio |
|---|---|---|---|---|
| 1% | Kalman LSE | 0.31 | 0.67 | 7 |
| 1% | SVGP-KAN LSE | 0.54 | 0.53 | 1 |
| 1% | SVGP-KAN SAMM | 0.60 | 0.54 | 1 |
| 5% | Kalman LSE | 0.24 | 0.64 | 8 |
| 5% | SVGP-KAN LSE | 0.78 | 0.57 | 1 |
| 5% | SVGP-KAN SAMM | 0.77 | 0.56 | 1 |
| 8.1% | Kalman LSE | 0.29 | 0.58 | 7 |
| 8.1% | SVGP-KAN LSE | 0.91 | 0.50 | 1 |
| 8.1% | SVGP-KAN SAMM | 0.92 | 0.52 | 1 |

*Note: Calibration slopes for SVGP-KAN methods exhibit moderate run-to-run variability (typically ±10–20%) due to stochastic optimization, while the qualitative findings remain consistent.*

The calibration slopes reveal a striking difference between Kalman and SVGP-KAN uncertainty. Kalman methods show slopes of around 0.2~0.3 across all sampling rates, indicating that predicted uncertainties are approximately 3–4 times larger than actual errors on average. Conversely, SVGP-KAN methods achieve slopes of around 0.5–0.9, substantially closer to the ideal value of 1.0.

However, the calibration slope alone does not tell the full story. The critical metric is the uncertainty ratio $\rho$, which measures whether the uncertainty correctly identifies when predictions are reasonably reliable. Kalman methods exhibit uncertainty ratios of 7–8, meaning the predicted uncertainty grows approximately 8 times between PIV observations relative to at PIV times. SVGP-KAN methods maintain ratios near 1.0, indicating that uncertainty remains proportional to actual prediction difficulty throughout the reconstruction.

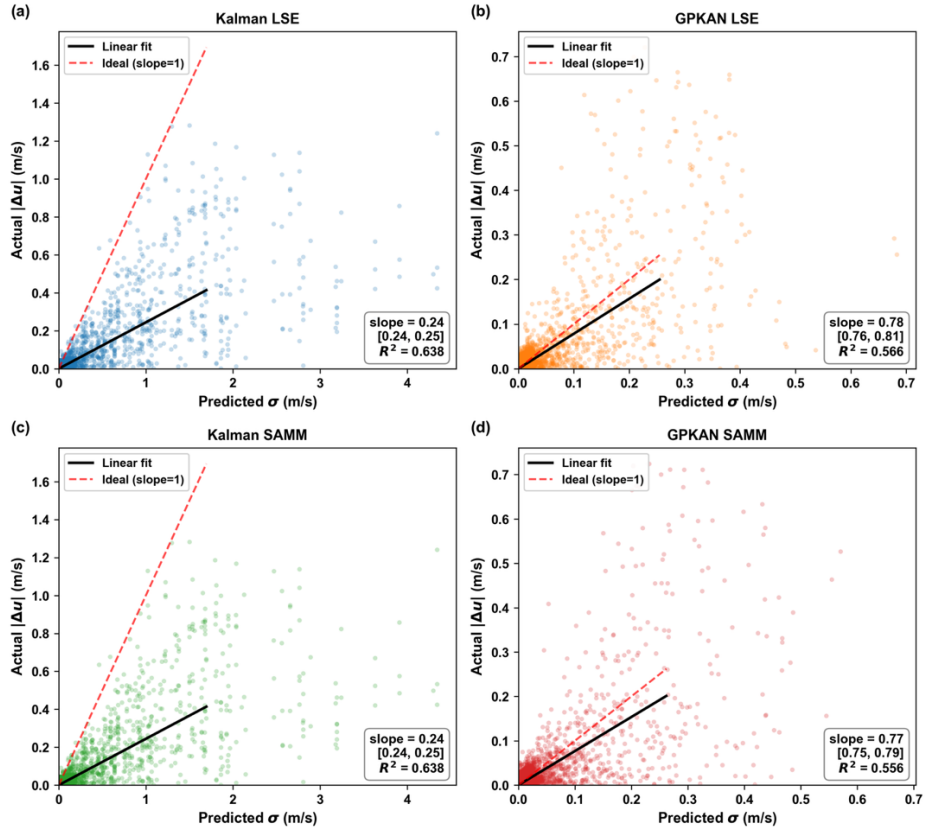

**Figure 9.** Uncertainty calibration curves at 5% sampling. Actual prediction error |Δu| versus predicted uncertainty σ for (a) Kalman LSE, (b) SVGP-KAN LSE, (c) Kalman SAMM, (d) SVGP-KAN SAMM. Dashed line indicates ideal calibration (slope = 1).

Figure 10 shows temporal uncertainty evolution at a representative spatial location for both SVGP-KAN and Kalman methods at 5% sampling. Green vertical lines mark PIV sample times. For SVGP-KAN LSE (left), the uncertainty bands (±1σ and ±2σ) remain relatively constant between PIV samples, with the reconstruction tracking the ground truth well. The uncertainty ratio is 1.3, indicating modest growth between observations.

For Kalman LSE (right), the uncertainty bands explode between PIV samples, growing approximately 8 times larger than at PIV times. While the wide bands contain the true values (achieving high coverage), they provide no useful information about prediction reliability. The same large uncertainty is reported regardless of whether the prediction is good or poor.

This visualization demonstrates the fundamental difference between Kalman covariance and GP-based uncertainty. Kalman covariance **P** measures uncertainty about the state given model assumptions. It grows according to the process noise covariance Q between observations and is reset by the measurement update, regardless of actual prediction difficulty. In contrast, SVGP-KAN uncertainty measures distance from training data in feature space, which naturally correlates with prediction difficulty. Regions and times far from training points have higher uncertainty.

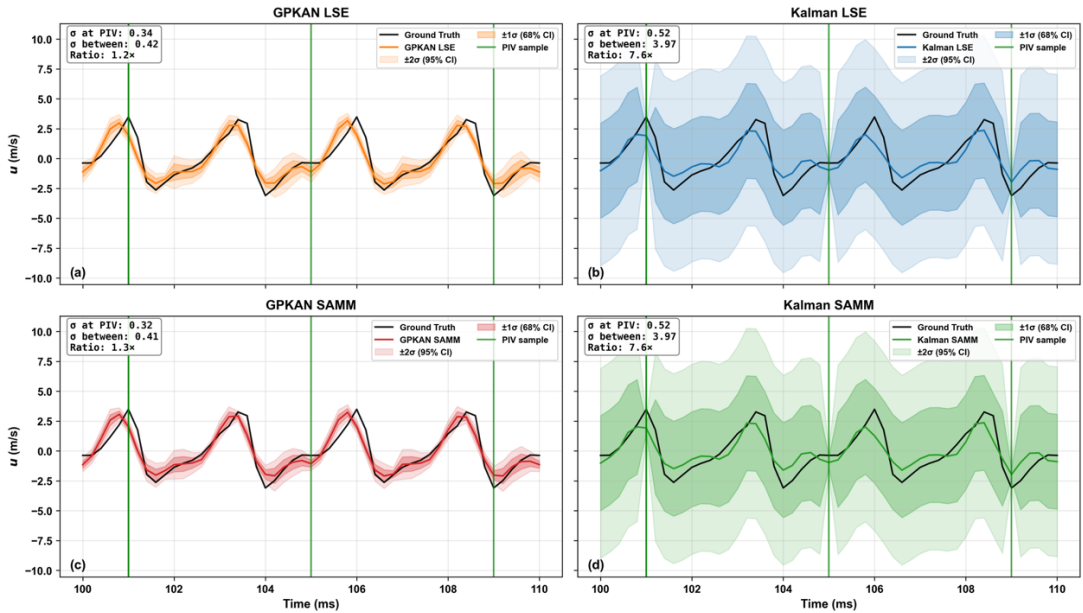

**Figure 10.** Temporal uncertainty comparison at 5% sampling showing reconstruction with uncertainty bands at a representative spatial location. (a) SVGP-KAN LSE (b) SVGP-KAN SAMM (c) Kalman LSE (d) Kalman SAMM.

Figure 11 shows the reliability diagram (coverage calibration) at 5% sampling. For each predicted confidence level, the observed coverage, which is defined as the fraction of true values falling within the predicted interval, is plotted. Perfect calibration corresponds to the diagonal.

Kalman methods appear slightly underconfident (above the diagonal), achieving high observed coverage at the 68% predicted level. However, this high coverage is achieved through uniformly wide intervals that provide no discrimination between reliable and unreliable predictions.

SVGP-KAN methods show moderate overconfidence (below the diagonal). Despite lower coverage, the SVGP-KAN uncertainty is *informative*: it correctly identifies times and locations where errors are larger.

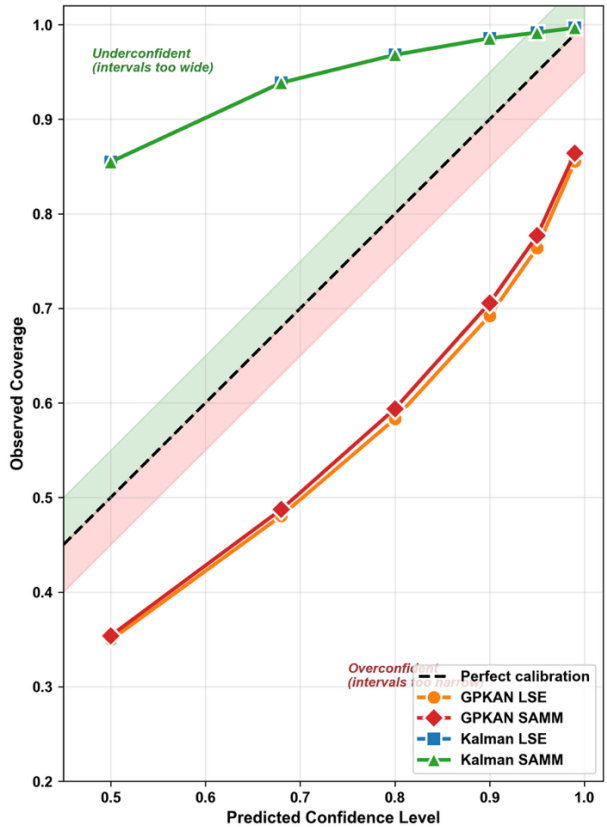

**Figure 11.** Reliability diagram at 5% sampling showing observed coverage versus predicted confidence level. Diagonal indicates perfect calibration.

## 3.9 Noise Sensitivity Analysis

To assess robustness under challenging measurement conditions, we conducted additional experiments at elevated noise levels (20% PIV noise, 2% sensor noise). Table 7 summarizes performance across noise conditions at 5% and 8.1% sampling.

Table 7: Performance comparison at low vs. high noise levels

| Sampling | Method | L2 (Low Noise) | L2 (High Noise) | Degradation |
|---|---|---|---|---|
| 5% | SVGP-KAN LSE | 0.054 | 0.068 | +26% |
| 5% | SAMM-RR | 0.054 | 0.069 | +28% |
| 5% | Kalman LSE | 0.112 | 0.123 | +10% |
| 8.1% | SVGP-KAN LSE | 0.047 | 0.059 | +26% |
| 8.1% | SAMM-RR | 0.047 | 0.059 | +26% |
| 8.1% | Kalman LSE | 0.087 | 0.096 | +10% |

We note that the method rankings preserved. SVGP-KAN and SAMM-RR maintain their accuracy advantage over Kalman methods even at elevated noise levels. All methods show modest degradation when noise is increased by 10 times, indicating robustness to measurement uncertainty. At high noise, SVGP-KAN calibration slopes decrease but the uncertainty ratio remains near 1.0, indicating that uncertainty still correctly tracks prediction difficulty.

## 3.10 Computational Cost

Table 8 summarizes computational requirements at 5% sampling rate. All results were obtained using a MacBook Air with M2 processor and 8 GB RAM without GPU acceleration. The SVGP-KAN library is currently unoptimized. SVGP-KAN methods require approximately 50 seconds total (training + inference), roughly 10–30X slower than classical methods. For offline analysis, this computational cost is acceptable given the uncertainty quantification capability. For real-time applications or rapid prototyping, SAMM-RR provides comparable accuracy with much faster computation, though without uncertainty estimates.

Table 8: Computational cost (5% sampling, 100 PIV snapshots)

| Method | Training + Inference Time |
|---|---|
| LSE (Vanilla) | ~5 s |
| LSE (Phase) | ~3 s |
| SAMM-RR | ~1.5 s |
| Kalman LSE | ~3 s |
| Kalman SAMM | ~3 s |
| SVGP-KAN LSE | ~50 s |
| SVGP-KAN SAMM | ~52 s |

## 4. Discussion

Our results suggest a decision framework based on available PIV sampling rate and application requirements (Table 9).

**Table 9: Method selection guidelines**

| Scenario | Recommended Method | Rationale |
|---|---|---|
| Accuracy critical, UQ not needed | SAMM-RR | Fast, accurate (L2 ≈ 0.047–0.054) |
| Accuracy + uncertainty needed | SVGP-KAN LSE | Best accuracy with calibrated UQ |
| Rapid prototyping | LSE (Vanilla) | Fastest, reasonable accuracy |
| Ultra-sparse (<2%), any scenario | Non-coprime + SVGP-KAN | Coprime fails and need samples/phase ≥2 |
| High noise (≥10% PIV) | SVGP-KAN with coprime ≥2 s/p | Maintains accuracy advantage |

For the present deterministic, coherent-structure-dominated flow, the globally linear SAMM-RR transfer function is adequate, explaining the comparable accuracy. For flows with stronger nonlinear mode interactions, such as turbulent flows with intermittent coherent structures or flows exhibiting bifurcations, SVGP-KAN's nonlinear modeling capability may provide advantages and warrants further systematic studies. More importantly, for any application requiring uncertainty quantification (design optimization under uncertainty, anomaly detection, confidence-weighted data assimilation), SVGP-KAN is preferred as it provides calibrated epistemic uncertainty that correctly identifies when predictions are less reliable.

The basic linear Kalman filter methods we implemented had two issues in this study. First, there are large generalization gaps between PIV and interpolation times, indicating that predictions degrade between observations. In addition, the Kalman covariance P fails to provide meaningful uncertainty estimates at interpolation times (calibration slope << 1 and uncertainty ratio >5). The global linear transition matrix $\mathbf{A}$, learned from sparse PIV-to-PIV transitions, cannot capture the flow over the long time-horizons ($\Delta t_{PIV} \approx 20$ sensor timesteps at 5% sampling) present in sparse sampling. This dynamics model mismatch leads to covariance inflation that fails to constrain the interpolation.

Several limitations of this study should be acknowledged. We implemented the forward Kalman filter and RTS smoother variants. Alternative dynamics models, such as nonlinear state-space models or physics-informed neural network dynamics, may address this limitation but were beyond the scope of this work.

The synthetic data represents an idealized case with perfectly periodic forcing and no cycle-to-cycle variability. Extension to turbulent flows would require: (1) higher POD truncation, (2) heteroscedastic GP formulations for spatially varying aleatoric uncertainty, and (3) potentially different sensor configurations. In turbulent flows, the epistemic uncertainty from SVGP-KAN would need to be distinguished from aleatoric uncertainty (turbulent fluctuations), requiring more sophisticated uncertainty decomposition.

The current configuration uses pressure measurements only. Temperature or other types of sensors could provide complementary information for thermal-fluid flows, potentially improving reconstruction accuracy through thermal-velocity correlations.

The SVGP-KAN computational cost (~50 s at 5% sampling) is acceptable for offline analysis but may limit real-time applications. GPU acceleration and mini-batch training could reduce this further. Altenative approaches to uncertainty quantification for neural networks, such as Bayesian neural networks, Monte Carlo dropouts, and ensemble Kalman inverstion, also warrants futher studies.

## 5. Conclusions

This work develops a machine learning approach to reconstruct flow fields using temporally sparse but spatially dense PIV data combined with spatially sparse but temporally dense sensor data. We integrate sparse variational Gaussian processes into Kolmogorov-Arnold Networks (SVGP-KAN), creating an architecture that provides both accurate reconstruction and calibrated epistemic uncertainty quantification.

Systematic evaluation on pulsed impingement jet flows reveals that SVGP-KAN, LSE, and SAMM-RR methods achieve comparable reconstruction accuracy. For this coherent, periodic flow, the globally linear SAMM-RR transfer function is adequate. The key advantage of SVGP-KAN lies in uncertainty quantification: the GP posterior correctly identifies when and where predictions are less reliable, maintaining calibration at interpolation times. For flows with stronger nonlinear mode interactions, SVGP-KAN's composed GP architecture may additionally provide accuracy advantages over linear methods.

We also find that a threshold of 2 samples per phase determines whether coprime or non-coprime sampling should be used. Above this threshold, coprime sampling achieves the best results through phase diversity; below it, non-coprime sampling is preferred for regression stability.

The SVGP-KAN framework provides a principled approach to uncertainty-aware flow reconstruction from sparse measurements, with applications extending beyond impingement jets to other periodic and quasi-periodic thermal-fluid systems.

## Acknowledgments

The author would like to thank Dr. Ukeiley for introducing the author to the foundational field reconstruction methods in his seminar.

## Declaration of generative AI and AI-assisted technologies in the manuscript preparation process

During the preparation of this work the author used Claude AI to help debug and refactor the python codes. After using this tool, the author reviewed and edited the content as needed and takes full responsibility for the content of the published article.

# References

[1] C. He, P. Wang, Y. Liu, Data assimilation for turbulent mean flow and scalar fields with anisotropic formulation, Exp. Fluids 62 (2021) 117. https://doi.org/10.1007/s00348-021-03213-8.

[2] S. Li, L. Ukeiley, Pressure-informed velocity estimation in a subsonic jet, Phys. Rev. Fluids 7 (2022) 014601. https://doi.org/10.1103/PhysRevFluids.7.014601.

[3] Y. Ozawa, T. Nagata, T. Nonomura, Spatiotemporal superresolution measurement based on POD and sparse regression applied to a supersonic jet measured by PIV and near-field microphone, J. Vis. 25 (2022) 1169–1187. https://doi.org/10.1007/s12650-022-00855-6.

[4] K.H. Manohar, O. Williams, R.J. Martinuzzi, C. Morton, Temporal super-resolution using smart sensors for turbulent separated flows, Exp. Fluids 64 (2023) 101. https://doi.org/10.1007/s00348-023-03639-2.

[5] S. Kaneko, A. del Pozo, H. Nishikori, Y. Ozawa, T. Nonomura, DMD-based spatiotemporal superresolution measurement of a supersonic jet using dual planar PIV and acoustic data, Exp. Fluids 65 (2024) 139. https://doi.org/10.1007/s00348-024-03872-3.

[6] B.C.Y. Yeung, O.T. Schmidt, Adaptive spectral proper orthogonal decomposition of tonal flows, Theor. Comput. Fluid Dyn. 38 (2024) 355–374. https://doi.org/10.1007/s00162-024-00695-0.

[7] Y. Kato, High-Resolution Visualization Measurement of Vortex-Shedding at High Frequencies Using Sub-Nyquist-Rate PIV and Compressed Sensing, J. Flow Vis. Image Process. 32 (2025). https://doi.org/10.1615/JFlowVisImageProc.2024054391.

[8] K. Taira, S.L. Brunton, S.T.M. Dawson, C.W. Rowley, T. Colonius, B.J. McKeon, O.T. Schmidt, S. Gordeyev, V. Theofilis, L.S. Ukeiley, Modal Analysis of Fluid Flows: An Overview, AIAA J. 55 (2017) 4013–4041. https://doi.org/10.2514/1.J056060.

[9] J. L. Lumley, The structure of inhomogeneous turbulent flows, in: Atmospheric Turbul. Radio Wave Propag., Nauka, 1967: pp. 166–178.

[10] G. Berkooz, P. Holmes, J.L. Lumley, The Proper Orthogonal Decomposition in the Analysis of Turbulent Flows, Annu. Rev. Fluid Mech. 25 (1993) 539–575. https://doi.org/10.1146/annurev.fl.25.010193.002543.

[11] L. Sirovich, Turbulence and the Dynamics of Coherent Structures Part I: Coherent Structures, Q. Appl. Math. 45 (1987) 561–571.

[12] P.J. Schmid, Dynamic mode decomposition of numerical and experimental data, J. Fluid Mech. 656 (2010) 5–28. https://doi.org/10.1017/S0022112010001217.

[13] J.H. Tu, C.W. Rowley, D.M. Luchtenburg, S.L. Brunton, J.N. Kutz, On dynamic mode decomposition: Theory and applications, J. Comput. Dyn. 1 (2014) 391–421. https://doi.org/10.3934/jcd.2014.1.391.

[14] O.T. Schmidt, T. Colonius, Guide to Spectral Proper Orthogonal Decomposition, AIAA J. 58 (2020) 1023–1033. https://doi.org/10.2514/1.J058809.

[15] K. Willcox, Unsteady flow sensing and estimation via the gappy proper orthogonal decomposition, Comput. Fluids 35 (2006) 208–226. https://doi.org/10.1016/j.compfluid.2004.11.006.

[16] B.R. Noack, P. Papas, P.A. Monkewitz, The need for a pressure-term representation in empirical Galerkin models of incompressible shear flows, J. Fluid Mech. 523 (2005) 339–365. https://doi.org/10.1017/S0022112004002149.

[17] G. Evensen, Sequential data assimilation with a nonlinear quasi-geostrophic model using Monte Carlo methods to forecast error statistics, J. Geophys. Res. Oceans 99 (1994) 10143–10162. https://doi.org/10.1029/94JC00572.

[18] R.J. Adrian, P. Moin, Stochastic estimation of organized turbulent structure: homogeneous shear flow, J. Fluid Mech. 190 (1988) 531–559. https://doi.org/10.1017/S0022112088001442.

[19] J.P. Bonnet, D.R. Cole, J. Delville, M.N. Glauser, L.S. Ukeiley, Stochastic estimation and proper orthogonal decomposition: Complementary techniques for identifying structure, Exp. Fluids 17 (1994) 307–314. https://doi.org/10.1007/BF01874409.

[20] C.E. Tinney, F. Coiffet, J. Delville, A.M. Hall, P. Jordan, M.N. Glauser, On spectral linear stochastic estimation, Exp. Fluids 41 (2006) 763–775. https://doi.org/10.1007/s00348-006-0199-5.

[21] Y. Zhang, L.N. Cattafesta, L. Ukeiley, Spectral analysis modal methods (SAMMs) using non-time-resolved PIV, Exp. Fluids 61 (2020) 226. https://doi.org/10.1007/s00348-020-03057-8.

[22] A. Towne, O.T. Schmidt, T. Colonius, Spectral proper orthogonal decomposition and its relationship to dynamic mode decomposition and resolvent analysis, J. Fluid Mech. 847 (2018) 821–867. https://doi.org/10.1017/jfm.2018.283.

[23] T. Suzuki, Reduced-order Kalman-filtered hybrid simulation combining particle tracking velocimetry and direct numerical simulation, J. Fluid Mech. 709 (2012) 249–288. https://doi.org/10.1017/jfm.2012.334.

[24] T. Suzuki, Y. Hasegawa, Estimation of turbulent channel flow at based on the wall measurement using a simple sequential approach, J. Fluid Mech. 830 (2017) 760–796. https://doi.org/10.1017/jfm.2017.580.

[25] C.E. Rasmussen, C.K.I. Williams, Gaussian Processes for Machine Learning, The MIT Press, 2005. https://doi.org/10.7551/mitpress/3206.001.0001.

[26] M. Titsias, Variational Learning of Inducing Variables in Sparse Gaussian Processes, in: Proc. Twelfth Int. Conf. Artif. Intell. Stat., PMLR, 2009: pp. 567–574. https://proceedings.mlr.press/v5/titsias09a.html.

[27] J. Hensman, N. Fusi, N.D. Lawrence, Gaussian processes for big data, in: Proc. Twenty-Ninth Conf. Uncertain. Artif. Intell., AUAI Press, Arlington, Virginia, USA, 2013: pp. 282–290.

[28] Z. Liu, Y. Wang, S. Vaidya, F. Ruehle, J. Halverson, M. Soljacic, T. Hou, M. Tegmark, KAN: Kolmogorov–Arnold Networks, Int. Conf. Represent. Learn. 2025 (2025) 70367–70413.

[29] S.A. Faroughi, F. Mostajeran, A.H. Mashhadzadeh, S. Faroughi, Scientific Machine Learning with Kolmogorov-Arnold Networks, (2025). https://doi.org/10.48550/arXiv.2507.22959.

[30] Z. Liu, P. Ma, Y. Wang, W. Matusik, M. Tegmark, KAN 2.0: Kolmogorov-Arnold Networks Meet Science, (2024). https://doi.org/10.48550/arXiv.2408.10205.

[31] Y.S. Ju, Scalable and Interpretable Scientific Discovery via Sparse Variational Gaussian Process Kolmogorov-Arnold Networks (SVGP KAN), (2025). https://doi.org/10.48550/arXiv.2512.00260.

[32] Y.S. Ju, Uncertainty Quantification for Scientific Machine Learning using Sparse Variational Gaussian Process Kolmogorov-Arnold Networks (SVGP KAN), (2025). https://doi.org/10.48550/arXiv.2512.05306.

[33] A. Damianou, N.D. Lawrence, Deep Gaussian Processes, in: Proc. Sixt. Int. Conf. Artif. Intell. Stat., PMLR, 2013: pp. 207–215. https://proceedings.mlr.press/v31/damianou13a.html (accessed December 22, 2025).

[34] R. Viskanta, Heat transfer to impinging isothermal gas and flame jets, Exp. Therm. Fluid Sci. 6 (1993) 111–134. https://doi.org/10.1016/0894-1777(93)90022-B.

[35] N. Zuckerman, N. Lior, Jet Impingement Heat Transfer: Physics, Correlations, and Numerical Modeling, in: G.A. Greene, J.P. Hartnett†, A. Bar-Cohen, Y.I. Cho (Eds.), Adv. Heat Transf., Elsevier, 2006: pp. 565–631. https://doi.org/10.1016/S0065-2717(06)39006-5.

[36] H.M. Hofmann, M. Kind, H. Martin, Measurements on steady state heat transfer and flow structure and new correlations for heat and mass transfer in submerged impinging jets, Int. J. Heat Mass Transf. 50 (2007) 3957–3965. https://doi.org/10.1016/j.ijheatmasstransfer.2007.01.023.

[37] H.M. Hofmann, R. Kaiser, M. Kind, H. Martin, Calculations of Steady and Pulsating Impinging Jets—An Assessment of 13 Widely used Turbulence Models, Numer. Heat Transf. Part B Fundam. 51 (2007) 565–583. https://doi.org/10.1080/10407790701227328.

[38] D. Simon, Optimal State Estimation: Kalman, H Infinity, and Nonlinear Approaches, John Wiley & Sons, 2006.

# Nomenclature

## Latin Symbols

| Symbol | Description | Units |
|---|---|---|
| $a_k$ | POD temporal coefficient for mode $k$ | — |
| **A** | Dynamics matrix (Kalman filter) | — |
| **C** | Transfer matrix (LSE) | — |
| $D$ | Nozzle diameter | m |
| $f$ | Frequency | Hz |
| $f_v, f_s, f_h$ | Varicose, sinuous, helical forcing frequencies | Hz |
| $\mathbf{G}_{yy}$ | Sensor cross-spectral density matrix | — |
| $\mathbf{G}_{ay}$ | Sensor-mode cross-spectral density | — |
| $H$ | Nozzle-to-plate distance | m |
| **H** | Transfer function (frequency domain) | — |
| $k$ | Steps between PIV observations | — |
| **K** | Kalman gain | — |
| $\mathbf{K}_{ZZ}$ | Kernel matrix at inducing points | — |
| $\ell$ | GP lengthscale | — |
| $M$ | Number of inducing points | — |
| $N_m$ | Number of POD modes | — |
| $N_{PIV}$ | Number of PIV snapshots | — |
| $N_s$ | Number of sensors | — |
| $N_x$ | Number of spatial grid points | — |
| **P** | Covariance matrix (Kalman filter) | — |
| **Q** | Process noise covariance | — |
| $r$ | Radial coordinate | m |
| **R** | Measurement noise covariance | — |
| $R_{dyn}$ | Dynamic range ratio | — |
| $Re$ | Reynolds number | — |
| $St$ | Strouhal number ($fD/U_m$) | — |
| $t$ | Time | s |
| $T$ | Period | s |
| $u$ | Radial velocity component | m/s |
| **u** | Velocity vector | m/s |

| Symbol | Description | Units |
|---|---|---|
| $U_m$ | Mean jet velocity | m/s |
| $w$ | Axial velocity component | m/s |
| $\mathbf{x}$ | Spatial coordinate vector | m |
| $\mathbf{y}$ | Sensor measurement vector | — |
| $z$ | Axial coordinate | m |
| $\mathbf{Z}$ | Inducing inputs (GP) | — |

## Greek Symbols

| Symbol | Description | Units |
|---|---|---|
| $\alpha$ | Calibration slope | — |
| $\Delta t$ | Timestep | s |
| $\epsilon$ | Observation noise | — |
| $\epsilon_{L2}$ | L2 relative error | — |
| $\kappa$ | Condition number | — |
| $\lambda$ | KL divergence weight | — |
| $\lambda_k$ | POD eigenvalue | — |
| $\mu$ | Predictive mean | — |
| $\phi$ | Forcing phase ($2\pi f_v t$) | rad |
| $\boldsymbol{\Phi}_k$ | POD spatial mode | — |
| $\varphi_{j,i}$ | KAN edge function | — |
| $\rho$ | Uncertainty ratio ($\sigma_{between}/\sigma_{PIV}$) | — |
| $\sigma^2$ | Variance | — |
| $\sigma_f^2$ | GP prior variance | — |
| $\omega$ | Vorticity | 1/s |

## Subscripts and Superscripts

| Symbol | Description |
|---|---|
| $(\cdot)^+$ | Moore-Penrose pseudoinverse |
| $(\cdot)^H$ | Conjugate transpose |
| $(\cdot)^-$ | Prior estimate (Kalman) |
| $(\cdot)'$ | Fluctuating component |
| $(\cdot)_{PIV}$ | At PIV measurement times |

| Symbol | Description |
| --- | --- |
| $(\cdot)_{rec}$ | Reconstructed |
| $(\cdot)_{true}$ | Ground truth |
| $\bar{(\cdot)}$ | Time-averaged |

## Abbreviations

| Abbreviation | Definition |
| --- | --- |
| CFD | Computational Fluid Dynamics |
| DMD | Dynamic Mode Decomposition |
| ELBO | Evidence Lower Bound |
| GP | Gaussian Process |
| KAN | Kolmogorov-Arnold Network |
| KL | Kullback-Leibler (divergence) |
| LSE | Linear Stochastic Estimation |
| PIV | Particle Image Velocimetry |
| POD | Proper Orthogonal Decomposition |
| PSD | Power Spectral Density |
| RBF | Radial Basis Function |
| SAMM | Spectral Analysis Modal Method |
| SAMM-RR | Spectral Analysis Modal Method –Rank Reduction |
| SLSE | Spectral Linear Stochastic Estimation |
| SVGP | Sparse Variational Gaussian Process |
| SVGP-KAN | Sparse Variational GP Kolmogorov-Arnold Network |
| TR-PIV | Time-Resolved PIV |
| UQ | Uncertainty Quantification |

## Data Availability Statement

The datasets and complete analysis code will be available upon reasonable requests to the author.

## Author Contributions

Conceptualization, methodology, software, validation, formal analysis, investigation, data curation, writing, and visualization: Y. S.J.

## Funding

This research received no external funding.

## Conflicts of Interest

The author declares no conflicts of interest.